\documentclass[a4paper,twoside]{article}

\usepackage{epsfig}
\usepackage{subcaption}
\usepackage{calc}
\usepackage{amssymb}
\usepackage{amstext}
\usepackage{amsmath}
\usepackage{amsthm}
\usepackage{multicol}
\usepackage{pslatex}
\usepackage{apalike}
\usepackage{algorithm2e}
\usepackage[bottom]{footmisc}

\usepackage{graphicx}   %
\graphicspath{{./figures/}}
\usepackage{float}
\usepackage{hyperref}
\usepackage{ragged2e}
\usepackage[capitalise,noabbrev]{cleveref}
\Crefname{figure}{Fig.}{Figs.}
\Crefname{tabular}{Tab.}{Tabs.}
\Crefname{section}{Sec.}{Secs.}
\usepackage{array}
\usepackage{colortbl}
\usepackage{xspace}

\newcommand{\eg}{e.\,g.}

\newcommand{\sysname}{CUVE\xspace} 

\definecolor{color_expert}{rgb}{0.106, 0.620, 0.467}
\definecolor{color_novice}{rgb}{0.851, 0.373, 0.008}
\newcommand{\expert}[1]{\textcolor{color_expert}{#1}}
\newcommand{\novice}[1]{\textcolor{color_novice}{#1}}
\newcommand{\errorbar}{Error bars represent 95\% bootstrapped CI. }

\usepackage{SCITEPRESS}     

\begin{document}

\title{CUVE: Making Supervised Classification\\ more Accessible through Visualization}

\author{\authorname{Sanchitha Kuppusami\sup{1}\orcidAuthor{0009-0005-9170-0516}, Xiyao Wang\sup{1}\orcidAuthor{0000-0003-2429-6064} and Miguel A. Nacenta\sup{1}\orcidAuthor{0000-0002-9864-9654}}
\affiliation{\sup{1}Department of Computer Science, University of Victoria, Victoria, Canada}
\email{sanchu.kuppusami@gmail.com, xiyao.wang23@gmail.com, nacenta@uvic.ca}
}


\abstract{We present CUVE, an interface that enables people to build classifiers by generating visualizations that are interpretable by both humans and learning algorithms (the Visualization as Intermediate Representation approach). This UI enables users to create supervised classifiers and to visually investigate how to improve them and how they carry out their function. A preliminary study investigates the usability of our prototype and compares it with a traditional text-based programming interface.}

\onecolumn \maketitle \normalsize \setcounter{footnote}{0} \vfill

\section{\uppercase{Introduction}}
\label{sec:intro}
An open problem in machine learning is that it is difficult for humans to understand and explain why a classifier has classified an instance in a particular way. This problem is often referred to as the ``black box'' problem of ML. It is important because the ability to inspect the mechanisms by which a decision is made is important for accountability, debugging and trust~\cite{doshi-velezRigorousScienceInterpretable2017}.

Several approaches have been proposed to ameliorate this issue within the subarea of \emph{explainable, interpretable or intelligible AI}. 
One proposed technique is using VisuaLization As Intermediate Representation (VLAIR~\cite{vlair,vlair_2}). The core idea is that, instead of training a deep learning network directly on data, it is trained on visualizations of the data. The transformation of the data into visualizations can be determined or guided by a human in analogously to how visualizations are designed for regular data: the specification of a mapping from data to visual marks and features. For example, a card transaction will generate a small visualization that could have a large red circle if the transaction comes from a location far from the credit card holder's home, and a small green circle if from a nearby location. Other features of the transaction will be mapped to other marks in the images, such as other shapes, points or gradients of different colors, sizes and lengths. The system is trained using these images as the labeled data. This technique partially addresses the black box problem because sharing a representation (the visualization) gives humans a visual understanding of the data and, importantly, gives the human an opportunity to infer visually how the ML classification works. For example, by inspecting the visualizations of instances classified as one category, a person can try to infer which visual features they have in common. In other words, the visual representation becomes a machine-human shared representation of the data that might allow the human to infer how the machine might be categorizing instances.

Current applications of VLAIR-like approaches allow its implementation by manually modifying the custom visualization and training software. E.g., Jiang et al.~\cite{vlair_2} use pyplot in a jupyter notebook to generate visualizations used to train a model with pytorch. This is powerful but only accessible to people with fluency in advanced tools and programming languages. Moreover, even for experts, it can be tedious and inefficient.

CUVE (Classification Understood through Visual Externalizations) is a tool that integrates visualization design and comparison and model training in a single UI to facilitate applying the VLAIR approach. This paper describes CUVE and a study's results comparing CUVE and a traditional notebook-based alternative. Non-experts had a more positive appreciation of the CUVE tool and found that it could improve several aspects of the classification implementation process for both experts and non-experts.

\section{Background and Related Work}
Our work concerns the task of \emph{supervised classification}, where a learning algorithm uses data labeled in advance, to train a model that can then assign a class to any item, based only on the value of its features~\cite{singhReviewSupervisedMachine2016}. 
Classification can take place on images or regular data. Computer Vision (CV) has long leveraged DL algorithms to classify images~\cite{byerlyCurrentStateArt2022}.

Jiang et al. proposed to leverage CV approaches to classify non-visual/non-spatial data, which they call VisuaLizations As Intermediate Representations (VLAIR~\cite{vlair,vlair_2}). The main idea is that any data can be transformed into a visualization in the same way that data analysis and visualization professionals visually chart their data~\cite{visiontothink,visbook}) and then, instead of training an algorithm directly on the raw, typically non-spatial data, the algorithm is trained with the generated visualizations. Jiang et al. showed accuracy advantages, which has also be used in other contexts for this reason~\cite{zengConvolutionalNeuralNetworks2014}. However, more interestingly than its performance, VLAIR enables both machine and humans to share a visual representation. This might be useful to ameliorate the ``black box'' problem~\cite{confalonieriHistoricalPerspectiveExplainable2021}, that is, the difficulty for humans to interpret or make sense of the process of classification.

The black box problem or, more generally, the issues of interpretability of complex machine learning models, are a longstanding and growing concern in the ML community and in society at large. There is a general acceptance that most ML models act as black box models because they are too complex for humans to understand how they make predictions~\cite{black_white}, at least without help. There are two slightly different aims in the literature that pursues this kind of facilitation. \emph{Interpretability} is mostly ``associated with the intuition behind the outputs of a model. The idea is that the more interpretable a ML system is, the easier it is to identify cause-and-effect relationships within the system’s inputs and outputs \cite{explainability_def}'' and \emph{Explainability} is ``associated with the internal logic and mechanics that are inside a machine learning system. The more explainable a model, the deeper the understanding that humans achieve in terms of the internal procedures that take place while the model is training or making decisions'' \cite{explainability_def}.

The overlapping research areas of AI interpretability and explainable AI (XAI) are too vast to survey here (examples of good surveys are \cite{peeking,explainableai,kalasampathLiteratureReviewApplications2025}). A popular classification of XAI methods distinguishes between \emph{model-specific} methods such as Class Activation Mapping and a family of related methods (CAM~\cite{zhou2016learning,selvaraju2016grad,selvaraju2017grad} (mainly used for CNN algorithms), backpropagation-based methods such as DeepLIFT~\cite{shrikumar2017learning}, Layer-wise Relevance Propagation~\cite{bach2015pixel}) and perturbation-based methods such as Occlusion~\cite{ivanovs2021perturbation} and RISE~\cite{petsiuk2018rise}; \emph{model-agnostic} methods, of which LIME~\cite{lime} and SHAP~\cite{shapley} are the best well known. Finally, a range of different methods have been used to represent the relationship between the model outcome and the features in the data, such as partial dependence plots~\cite{pdp_ice}, ice~\cite{ice} and ale~\cite{ale}. Key differences between these types of visualizations and VLAIR is that in VLAIR the human designs the visualization, and the visualization is an output to be consumed both by the human and the machine.

\section{Goals and Design Principles}
\label{sec:design_goals}

We have two main goals; we seek to make ML classification more interpretable (G1) and we seek to make ML classification more accessible (G2). We address G1 by implementing a pure VLAIR approach, and G2 by providing an interface that does not require textual programming.

To address the goals, we adopted a set of design principles: \textbf{DP1} \emph{Support operation without coding} (avoid writing textual programming whenever possible by leveraging the constructive visualization approach~\cite{constructive_approach}); \textbf{DP2} \emph{Leverage visuals as much as possible} (the design should use visuals and visualizations as much as possible); \textbf{DP3} \emph{Enable iterative refinement} (support progressive iterative refinement of the model and of the users understanding of the dataset and how the model operates); \textbf{DP4} \emph{Avoid unnecessary complexity} (the design should be clean, simple, include only necessary elements and use familiar UI language); \textbf{DP5}  \emph{Avoid menus and hidden operations} (expose as much functionality as possible without requiring complex navigation); and \textbf{DP6} \emph{Avoid reliance on visual memory} (Display relevant information on one single screen when possible). 

\section{\sysname: Design}
\label{sec:design_sys}
In this section we describe CUVE's design, linking its features to the goals and design principles.

CUVE is implemented as a web application designed around five steps (web tabs), each corresponding to a step of the supervised classification workflow. It does not require coding because functions are available through widgets and canvases (\textbf{DP1}, \textbf{DP2}). Each tab/step leads to the next, creating a new tab in the browser. Opening a new tab does not close the previous one so that the user can switch back, make changes, and compare without memorizing or reconstructing everything (\textbf{DP3}). 

\begin{figure}[h]
         \centering
         \includegraphics[width=\columnwidth]{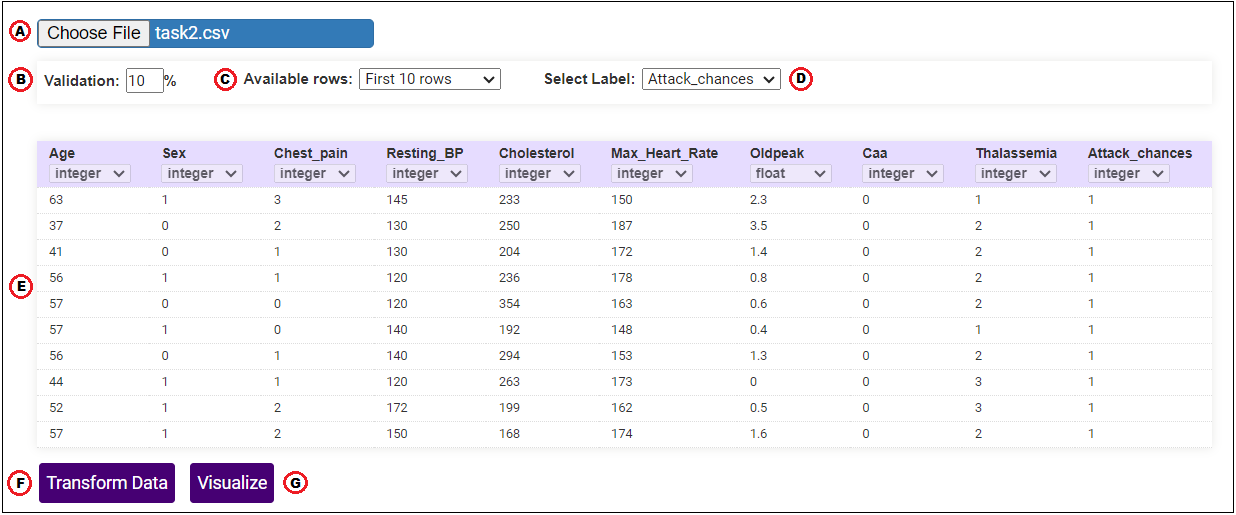}
         \caption{File Uploading and Inspection screen.}
         \label{fig:ivlair_fileupload}
 \end{figure}

Tab 1 displays a big data table representation (Figure~\ref{fig:ivlair_fileupload}) which shows the result of uploading data in .xlsx and .csv formats, and that Tab 2 enables selection of the data types for each column, addressing missing data situations, and creating new columns with simple formulas based on other columns.

\begin{figure*}[h]
         \centering
         \includegraphics[width=1.7\columnwidth]{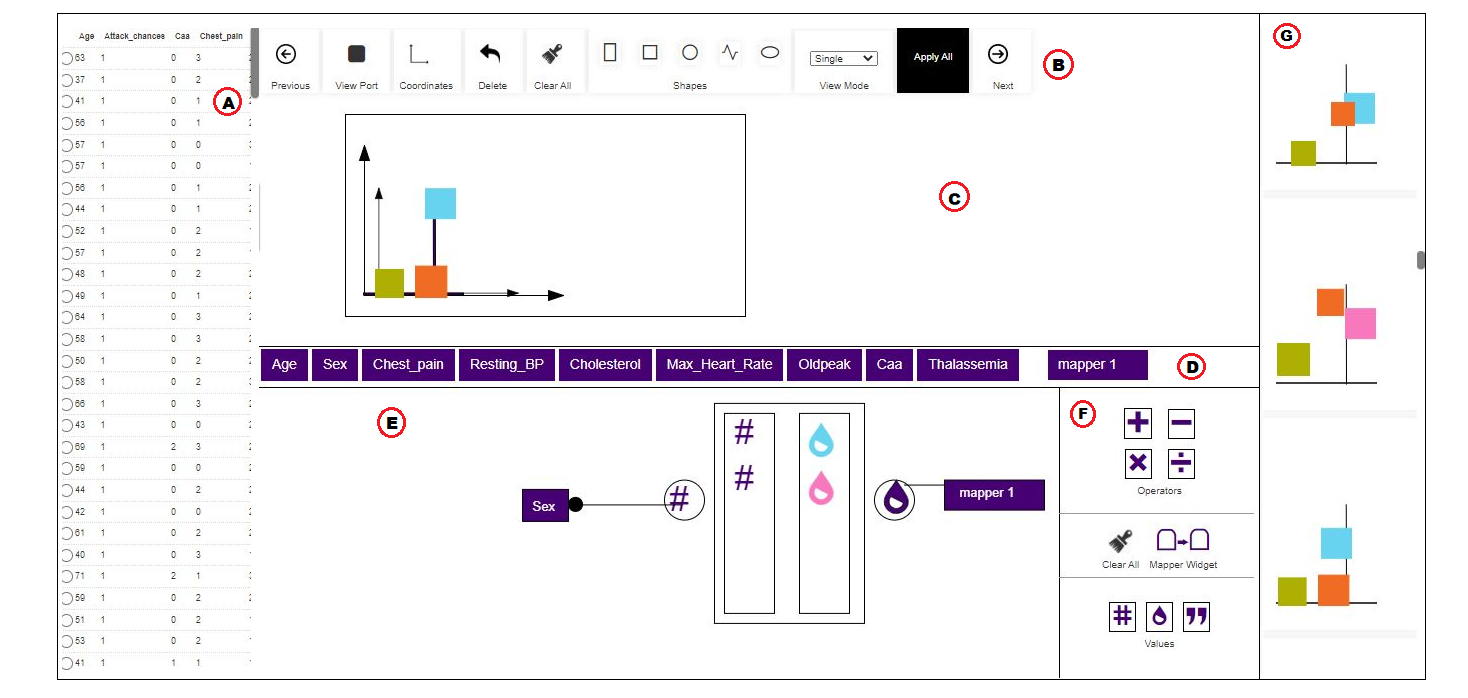}
         \caption{Tab 3: Creating visualizations.}
         \label{fig:ivlair_vis}
 \end{figure*}

\label{sec:screen3}
Tab 3 (\cref{fig:ivlair_vis}) corresponds to input selection and is also an extension of feature engineering in the traditional ML workflow. It lets users select and combine desired features to construct a visualization, which will later serve as input for classification. It is based on iVolver~\cite{ivolver}, a visual constructive visualization approach which supports graphical manipulation, accommodating design goals \textbf{DP4} and \textbf{DP5} (this approach has also been used in other contexts~\cite{halaharvi_p-inti_2024,mendez_bottom-up_2017,mendez_considering_2018}). It supports multiple functions.

In Tab 3 users create marks (e.g., rectangles, and circles) by dragging and dropping from the top panel (Figure~\ref{fig:ivlair_vis}.B) into the construction canvas (Figure~\ref{fig:ivlair_vis}.C). which also makes a dedicated panel appears to let users define the mark's visual parameters, such as size and color. These parameters can be linked to data features displayed as buttons in the middle panel (Figure~\ref{fig:ivlair_vis}.D).

The system has a default 2D coordinate system on the canvas, allowing users to change each mark's position. Each mark's orientation, and scale can also be changed or linked to data. 
Additionally, users can create other coordinate systems with relative transformations to the default one. The position of coordinate systems can also be associated to data variables.

Tab3 offers a transformation Canvas (Figure~\ref{fig:ivlair_vis}.D) that can combine and transform data columns into outputs that, in turn, can feed the positions or attributes of marks.

Finally, the user can preview the visualizations created for each row of the data. The user first selects which section of the visualization will fit within the output images. Then, selecting data rows in the left panel (Figure~\ref{fig:ivlair_vis}.F) displays the corresponding visualizations in the right panel (Figure~\ref{fig:ivlair_vis}.G). It is also possible to visualize a superposition of all the outputs to see the ranges and density of outputs in different regions of the image. 
 
\begin{figure}[h]
         \centering
         \includegraphics[width=\columnwidth]{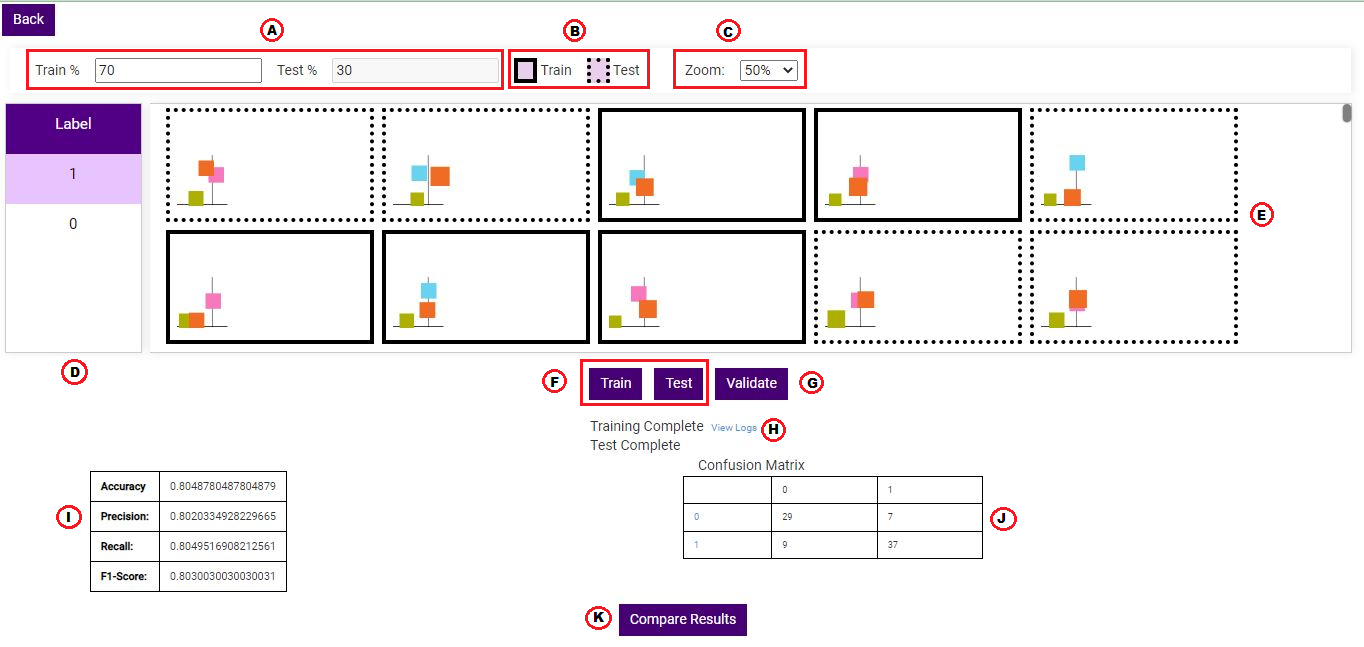}
         \caption{Algorithm training and tests.}
         \label{fig:ivlair_train}
 \end{figure}

\label{sec:screen4}
Tab 4 (\cref{fig:ivlair_train}) supports training and testing. Users visually inspect produced images, define and adjust the training and test set ratio, train a model, test it, overview  the results, and validate the model.

Upon entering tab 4, all images created from the previous tab are first grouped by labels, i.e., by the assigned classification output (\cref{fig:ivlair_train}.A). For each label, images are displayed (\cref{fig:ivlair_train}.E) and randomly separated into training and test datasets with a user-definable ratio (\cref{fig:ivlair_train}.A). Their role (training or test) is visually distinguished by using solid (training) and dotted (test) outlines (\textbf{DP2}). Users can manually change whether an image is selected for training or test by clicking on it.

When pressing on the ``train'' button (\cref{fig:ivlair_train}.F), the system starts to train the model with the provided dataset in the background. The system can display training logs and will show a message when it is finished to inform the user (\textbf{DP5}).
After training, users can press the test button and test (\cref{fig:ivlair_train}.F). The system then shows a confusion matrix of true and false positive and negatives (\cref{fig:ivlair_train}.J), as well as four selected evaluation metrics (accuracy, precision, recall, and F1-score, \cref{fig:ivlair_train}.I). In-depth result analysis happens in the next tab. 
When users are satisfied with the test results and done experimenting with different visualization mappings in other tabs, they can perform the validation with a portion of data (this is configurable in Tab 1). The four metrics and the confusion matrix also update with the validation results. Performing validation means the end of all operations, thus users should no longer modify any tabs.

\begin{figure}[h]
         \centering
         \includegraphics[width=\columnwidth]{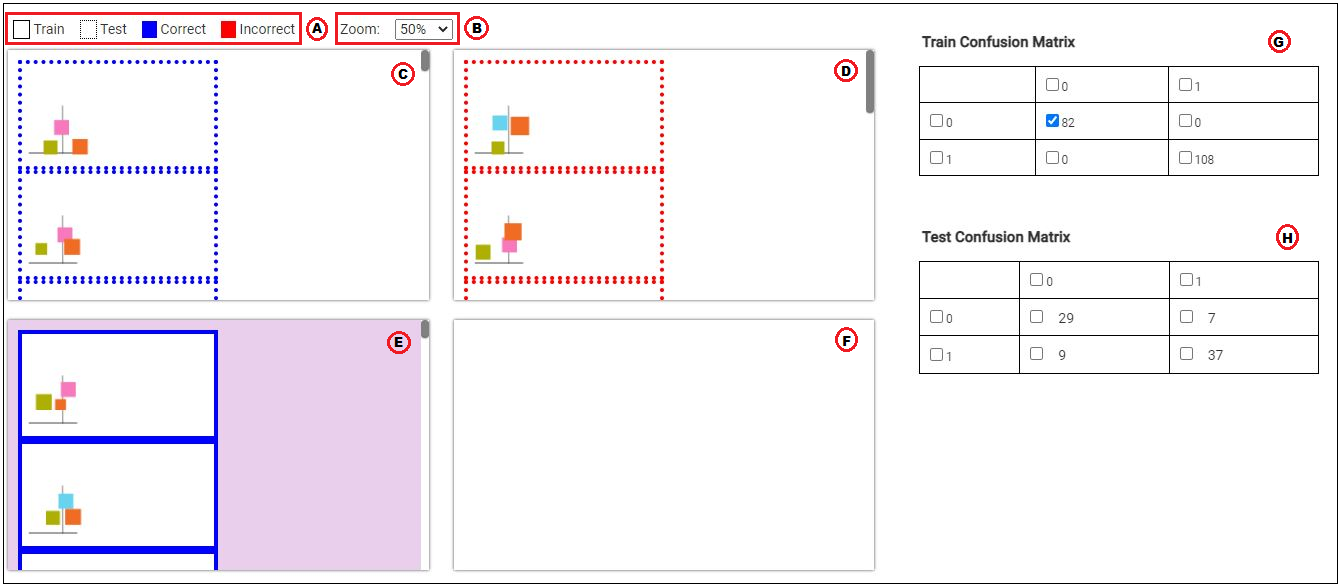}
         \caption{Algorithm training and tests.}
         \label{fig:ivlair_compare}
\end{figure}

\label{sec:screen5}
Tab 5 (\cref{fig:ivlair_compare}) allows users to group the images according to their classification results and visually inspect them to identify patterns. The goal is to make-sense from the results and generate insights for improvements. 
This tab displays two confusion matrices for training and testing (marked G and H in \cref{fig:ivlair_compare} respectively) and four panels (C, D, E, and F) for image visualization. Each value on the confusion matrix is associated with a checkbox; enabling a checkbox will display all images of this group on the selected panel. All panels can thus be configured separately and flexibly adjusted to help users understand the results through comparison (\textbf{DP4}).

This design is intended to facilitate the comparison of different groups of results visually. Users visually assess the differences and similarities between cases which were correctly and incorrectly classified, perhaps allowing the user to improve the visual mapping for better classification or to avoid errors. Visual comparisons between the categories can reveal the patterns that make them different. 

\section{\sysname: Evaluation}
\label{sec:eval}
We conducted a user study (IRB-approved) where participants developed an ML model with both \sysname and Python during a single 60-75 min session, for \$15. We recruited 16 participants (6 female and 10 male, aged between 22 and 39) in two groups: \novice{non-experts} only knew what ML is but did not have practical experience, while \expert{experts} had created at least one ML model by themselves. All at least basic knowledge of Python programming and ML.

\subsection{Study Design, Tasks and Procedure}
\label{sec:eval_design}\label{sec:eval_procedure}
All participants completed two phases for each of two conditions (C1:VLAIR and a C2:Jupyter Notebook). 
 
The order of conditions was counterbalanced; participants used a different dataset in each condition. 

\begin{figure}[h]
         \centering
         \includegraphics[width=\columnwidth]{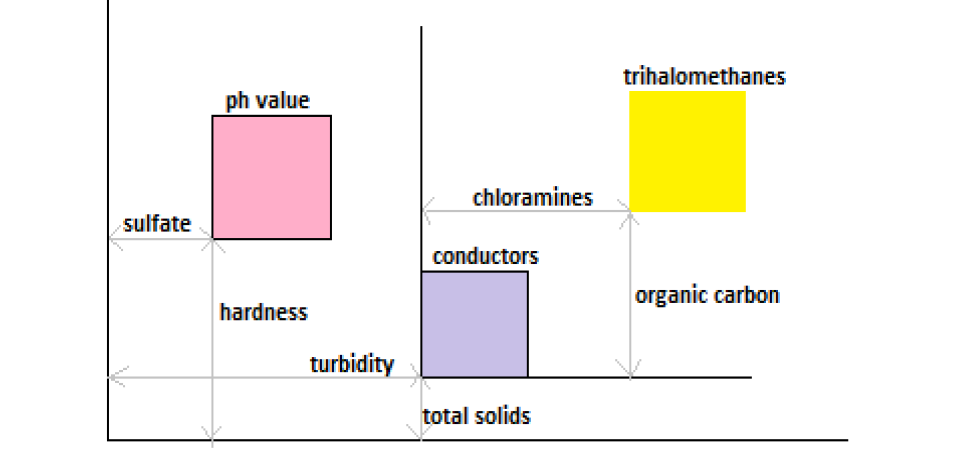}
         \caption{An example of the Visualization that the instructor guided the participate to create during the study, using water probability datasets.}
         \label{fig:example_vis}
 \end{figure}

In Phase 1 the experimenter guided the participant through a multi-step supervised classification process. For the \sysname condition, we pre-loaded the dataset. Then, tab by tab, we asked them what they expected was the functionality of each tab, and then taught them to use it. For constructing the visualization, we guided them step by step to create an example visualization mapping (\cref{fig:example_vis}). For the Python condition we provided the Python code in four cells: data loading, feature engineering, model training and classification, and results display. A set of functions allowed participants to quickly modify features, including deleting, creating new features based on existing ones, and performing arithmetic operations and to query the results. In Phase 2 participants played freely with each system and modified it. After each condition they filled a questionnaire. At the end, asked follow-up questions about their preferences and perceived differences between the two tools.

\section{Results}
We gathered and analyzed both quantitative data (SUS and preference ratings)and qualitative data (comments, feedback, and answers to the interview questions) from each of the 16 participants. 

\subsection{Quantitative Analysis}

\begin{figure}
         \centering
         \includegraphics[width=\columnwidth]{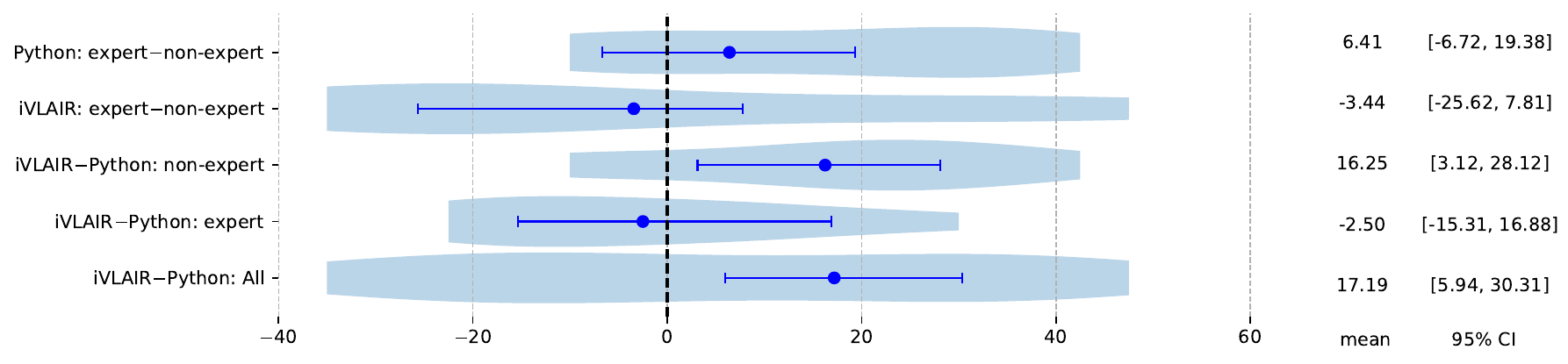}
         \caption{Average differences in aggregated SUS scores. \errorbar}
         \label{fig:difference_overallscore}
 \end{figure}

Non-experts rated \sysname with a conclusively higher SUS score than Python (on average, a difference of 16.3---see Figure~\ref{fig:difference_overallscore}). 
When considering all participants together, the difference is only 6.4 on average.
Experts also generally rated Python higher than non-experts (17.1 avg. difference).

4 out of 8 experts preferred \textbf{\sysname}. For non-experts 6 preferred the Python condition.

\subsection{Qualitative Analysis}
We combined inductive and deductive coding with thematic analysis to analyze participants' comments. The authors iterated on a codebook that, in its final version, contained 45 codes and 5 themes. Here we only summarize the most salient results.

All participants mentioned that \textbf{\sysname} was helpful to understand the data and/or to find patterns when looking at the small multiples (\eg, \emph{``because like even when I was seeing for 20 seconds, I started seeking like a pattern ...''} [\novice{P1}], \emph{``it's easier to see when I have visuals to compare more quickly than just numbers} [\expert{P12}]).  

Three experts (\expert{P2}, \expert{P5}, and \expert{P8}) and three non-experts (\novice{P1}, \novice{P6}, \novice{P13}) said that it was easy to understand why a particular prediction is made and to debug the model (\eg, \emph{``I think that one could be used when you are stuck and you don't know what is wrong with your model''} [\novice{P8}].

When using \textbf{Python}, twelve participants complained about the difficulty of understanding or finding patterns in data. For example, \novice{P1} said that \emph{``I would need time to look at them because it's hard to figure out if there is a similarity because I need to go through the numbers at least one by one''}. \expert{P5} specifically mentioned that he could not understand why the model made that prediction (\emph{``I can't understand why the false positive values have occurred here''}). In contrast, four other participants (\expert{P4}, \expert{P7}, \expert{P8}, and \expert{P10}) who were all ML experts said that they were able to find at least some patterns in data  (\emph{``Not much. But yes, It does''} [\expert{P10}]).

7 out of 16 participants (\expert{P2}, \expert{P5}, \novice{P6}, \expert{P7}, \expert{P11}, \expert{P12}, and \novice{P15}) confirmed that \textbf{\sysname} helped them learn/understand ML without coding \emph{``I think it would benefit new users or new people that are learning machine learning''} [\expert{P12}],\emph{``help you visualize the data like easily and you do not really have to know how to code. ... It has a lot of built-in features that you can use easily without programming...'' }[\expert{P11}]). Nevertheless, several participants (including \novice{P1}, \novice{P3}, \expert{P4}, \expert{P5}, \novice{P6}, \expert{P7}, and \expert{P8}), insisted that some prior knowledge is still required.

Reusability came up as an issue of \sysname compared to the Python approach because they have to create a new visualization for every problem (\novice{P15}) whereas they don’t need to do it with \textbf{Python} (\expert{P12} and \novice{P15}).

\section{Discussion, Limitations and Conclusion}
Overall we have demonstrated, through the implementation of \sysname, the feasibility of a VLAIR approach as an end-user interactive system. \sysname also incorporates several small innovations: it allows users to construct visualizations visually and it offers comprehensive features to compare the small multiples visualizations generated through the visual mapping. 

An initial evaluation showed that the approach could be viable for both experts and non-experts as a UI that does not require programming. Participants appreciate the improved ability to understand the data that the visual approach offers. However, participants remain skeptical of replacing programming completely and suspect that the non-programming approach might also have additional demands such as visualization expertise.



\bibliographystyle{apalike}
{\small
\bibliography{bibliography.bib}}



\end{document}